\documentclass[prc,aps,superscriptaddress,floatfix,twocolumn,nofootinbib]{revtex4-2}
\usepackage{tipa}
\usepackage{float}
\usepackage{color}
\usepackage[colorlinks,
			linkcolor=blue,
			anchorcolor=blue,
			citecolor=blue]{hyperref}
\usepackage{amsmath}
\definecolor{gray}{RGB}{220,220,220}

\usepackage{graphicx}
\usepackage{dcolumn}
\usepackage{bm}
\usepackage{comment}
\usepackage{subfigure}

\def\beq{\begin{equation}}
	\def\eeq{\end{equation}}
\def\bea{\begin{eqnarray}}
	\def\eea{\end{eqnarray}}

\def\fun#1#2{\lower3.6pt\vbox{\baselineskip0pt\lineskip.9pt
		\ialign{$\mathsurround=0pt#1\hfil##\hfil$\crcr#2\crcr$\sim$\crcr}}}
\preprint{}
\begin{document}

\title{Impact of neutron–proton pairing on the nucleon high-momentum distribution in symmetric nuclear matter
}

\author{Guo-peng Li}
 \affiliation{Institute
of Modern Physics, Chinese Academy of Sciences, Lanzhou 730000, China}
\affiliation{ School of Nuclear Science and Technology, University of Chinese Academy of Sciences, Beijing 100049, China}
\author{Ji-you Fu}
\affiliation{Institute
of Modern Physics, Chinese Academy of Sciences, Lanzhou 730000, China}
\affiliation{ Institute of Theoretical Physics \& Research Center of Gravitation, Lanzhou
 University, Lanzhou 730000, China}
\author{Jin Zhou}
 \affiliation{Institute
of Modern Physics, Chinese Academy of Sciences, Lanzhou 730000, China}
\affiliation{ School of Nuclear Science and Technology, University of Chinese Academy of Sciences, Beijing 100049, China}
\author{Xin-le Shang}\email[ ]{shangxinle@impcas.ac.cn}
\affiliation{Institute
of Modern Physics, Chinese Academy of Sciences, Lanzhou 730000, China}
\affiliation{ School of Nuclear Science and Technology, University of Chinese Academy of Sciences, Beijing 100049, China}
\author{Jian-min Dong}
\affiliation{Institute
of Modern Physics, Chinese Academy of Sciences, Lanzhou 730000, China}
\affiliation{ School of Nuclear Science and Technology, University of Chinese Academy of Sciences, Beijing 100049, China}
\author{Wei Zuo}
\affiliation{Institute
of Modern Physics, Chinese Academy of Sciences, Lanzhou 730000, China}
\affiliation{ School of Nuclear Science and Technology, University of Chinese Academy of Sciences, Beijing 100049, China}

\begin{abstract}
The effect of neutron-proton ($np$) pairing on the high-momentum tail (HMT) of nucleon momentum distributions in symmetric nuclear matter is investigated within a combined framework of the extended Brueckner–Hartree–Fock approach with off-shell BCS theory. The HMT ratio, quantifying the high-momentum fraction in the BCS state relative to the normal state, reaches about $1.06$ around the density of $0.052\ \mathrm{fm}^{-3}$, indicating that the maximal contribution of the $np$ pairing, amounts to approximately 6\% that from short-range correlations (SRCs). This contribution exhibits a density dependence that closely follows the squared relative pairing gap $\widetilde{\Delta}_F=Z_F\Delta(k_F)$ with respect to the kinetic energy $E_{k_F}^*$ evaluated using the effective mass, suggesting that $\widetilde{\Delta}_F^2/E_{k_F}^{*2}$ provides a qualitative measure of the $np$ pairing effect on the HMT. These findings highlight the significant role of $np$ pairing and its interplay with SRCs in shaping nucleon momentum distributions in nuclear matter.

\end{abstract}

\maketitle

\section{Introduction}
The nucleon momentum distribution plays a crucial role in revealing the underlying correlations and dynamics within the nucleus~\cite{review1,sci2008,15HM,LZH}. In an ideal, non-interacting Fermi gas at zero temperature, the momentum distribution follows a step function $n(k)=\theta(k_F-k)$, indicating full (null) occupation of states below (above) the Fermi momentum. In contrast, realistic nucleon-nucleon ($NN$) interactions introduce correlations that deplete the Fermi sea and populate momentum states above the Fermi surface~\cite{Fermisea,SHANG2020}. This depletion strongly affects the transport properties of nuclear matter, as well as related astrophysical phenomena in neutron stars, such as cooling~\cite{Dong_2016}, gravitational-wave-driven $r$-mode instability~\cite{SHANG2020}, and glitch phenomena~\cite{Shang2021}, by modifying the quasiparticle strength near the Fermi surface. Meanwhile, the emergence of a high-momentum tail in the momentum distribution provides direct evidence of short-range correlations in momentum space~\cite{sci2008,sci2014,2-2.8,ZXYang}. These HMTs originate from strong short-range $NN$ interactions between nucleons, leading to pronounced deviations from the independent-particle motion of the nucleus~\cite{sci2008,independent,Deshalit1974,Bohr1988}.
Furthermore, the presence of an HMT can significantly influence the production of $\pi^+$ mesons and hard photons in intermediate-energy heavy-ion collisions~\cite{YONG2017}, where hard photons, in particular, serve as sensitive probes of the early-stage momentum distribution in the collision zone~\cite{15HM,photons}. Recent precise measurement of bremsstrahlung $\gamma$-ray emission in low-energy heavy-ion collisions has provided unambiguous statistical evidence for SRC-induced HMT components in nuclei~\cite{XZG2025,XZG2026}. Therefore, a comprehensive understanding of the microscopic origin and quantitative features of HMT in nuclear matter is essential to gain a deeper understanding of the fundamental aspects of $NN$ correlations and interactions.

At short distances, the strong repulsive core of the $NN$ interaction gives rise to hard-core correlations that generate pronounced high-momentum components above approximately 600 MeV/c~\cite{2-2.8}. In addition to this short-range repulsion, tensor correlations, originating from the tensor component of the $NN$ potential, also play an essential role in shaping the HMT of the nucleon momentum distribution. The tensor force acts predominantly in spin-triplet, isospin-singlet ($S=1$, $T=0$) neutron–proton ($np$) pairs~\cite{sci2014,S=0-T=1Tensor}, which constitute the major source of SRCs. Experimental measurements from two-nucleon knockout reactions~\cite{review1,eep1,eep2}, as well as recent hard-photon measurements in low-energy heavy-ion collisions~\cite{XZG2025,XZG2026}, show that about 20\% of nucleons in nuclei can form short-range correlated pairs with large relative momenta and small center-of-mass momenta~\cite{center-of-mass}. These correlated pairs correspond to the occupation of high-momentum states. Among these correlated pairs, the number of $np$ pairs is found to be approximately 18 times that of proton-proton ($pp$) pairs~\cite{sci2008}, confirming the dominant contribution of the tensor correlations to the formation of the HMT. Although theoretical studies suggest that hard-core repulsion becomes dominant at higher momenta ($k>3 \ \mathrm{fm}^{-1}$), whereas the tensor force approximately contributes to the intermediate-momentum region ($k\approx 2-2.8\ \mathrm{fm}^{-1}$)~\cite{2-2.8,LYU}, both theoretical calculations and experimental measurements continue to face challenges in unambiguously distinguishing the regions dominated by hard-core and tensor correlations.

One should note that, besides the hard-core and tensor correlations that generate HMT, neutron–proton pairing (i.e., isoscalar pairing) arising from the strong long-range attraction of the tensor force in symmetric nuclear matter can also induce a high-momentum distribution~\cite{shang2013,duan2025}. From the occupation probability of particles in the pairing state, $n_{BCS}(k)=1-(k^2/2m-\mu)/\sqrt{(k^2/2m-\mu)^2+\Delta^2}$, it is straightforward to obtain the asymptotic behavior at large momenta, $n_{BCS}(k)\propto C/k^4$. This behavior is, to some extent, similar to the Tan relation~\cite{tan1,tan2,tan3}, a simple formula widely used to describe HMTs~\cite{Hen,PhysRevLett.104.235301}. This implies that $np$ pairing may also contribute to the high-momentum distribution of nucleons in nuclear matter. A natural and interesting question then arises: to what extent do $np$ pairing and SRCs collectively influence the high-momentum tail?  
Quantifying this effect not only helps one to elucidate the microscopic connection between pairing and SRCs but also deepens our understanding of the tensor force, particularly its role at long ranges, and the nucleon–nucleon (NN) interaction.

To address this issue, we employ a consistent theoretical framework combining the extended Brueckner–Hartree–Fock (EBHF) approach~\cite{EBHF1,EBHF3,EBHF4,EBHF5} with off-shell BCS theory, thereby incorporating both SRCs and pairing effects. In this framework, the EBHF approach provides the nucleon self-energy, which accounts for the SRCs through the ladder diagram, with the Pauli operator preventing the intermediate nucleons from scattering into the Fermi sea~\cite{shang2021BBG}. Although the EBHF self-energy captures the effects of SRCs, it cannot incorporate pairing effects, as the corresponding $G$-matrix associated with the ladder diagram may develop spurious singularities~\cite{han2023} that essentially arise from $np$ pairing~\cite{singular1,singular2,singular3}. By contrast, the off-shell BCS calculation based on the self-energy obtained can naturally account for both SRCs and the pairing effects~\cite{fxh,BOZEK200393}. Within this combined framework, we are able to consistently evaluate nucleon momentum distributions, spectral functions, and pairing gaps.



The paper is organized as follows: in Sec. II, we briefly review the theoretical methods employed, including the EBHF approach, the off-shell BCS gap formalism, and the corresponding spectral functions and momentum distributions. The results and discussion are presented in Sec. III, followed by a summary in Sec. IV.

\section{Formalism}
In the present calculation, the nucleon self-energy $\Sigma(k,\omega)$ is obtained within the EBHF framework using the effective interaction matrix $G$ after several self-consistent iterations. Details of the formalism can be found in Refs.~\cite{zuo1999,shang2021BBG}. The self-energy can be expanded in a perturbation series according to the number of hole lines~\cite{holeline}. The lowest-order term corresponds to the BHF approximation and can be written as
\begin{align}
    M_1(k,\omega)=\sum\limits_{k^{\prime}}n(k^{\prime})\langle kk^{\prime}|G[\omega+\epsilon(k^{\prime})]|kk^{\prime}\rangle_A,
\end{align}
where $\omega$ is the starting energy, $\epsilon(k)$ denotes the single-particle energy spectrum in the BHF approximation, and $n(k)$ reduces to the step function $\theta(k-k_F)$ at zero temperature. The subscript $A$ indicates antisymmetrization of the matrix element.

The second-order term in the hole-line expansion, referred to as the \emph{rearrangement} term, is expressed as
\begin{align}
M_2(k,\omega) & =\frac{1}{2}\sum\limits_{k^{\prime}k_1k_2}[1-n(k^{\prime})]n(k_1)n(k_2) \nonumber\\
 & \times\frac{|\langle kk^{\prime}|G[\epsilon(k_1)+\epsilon(k_2)]|k_1k_2\rangle_A|^2}{\omega+\epsilon(k^{\prime})-\epsilon(k_1)-\epsilon(k_2)-i0},
\end{align}
which accounts for particle-hole excitations in nuclear matter~\cite{M2}. 

In symmetric nuclear matter, the strong tensor correlations require the inclusion of the third-order contribution to describe the $np$ pairing~\cite{fxh}. This so-called \emph{renormalization} term~\cite{M2,M3}, which actually accounts for the partial occupation of hole states below the Fermi surface due to the $NN$ correlations, is given by
\begin{align}
    M_3(k,\omega)=-\sum\limits_{h^\prime}\kappa_2(h^\prime)\langle kh^\prime|G[\omega+\epsilon(k^\prime)]|kh^\prime\rangle_A,
\end{align}
where $\left. \kappa_2(h^{\prime})=-\left[\frac{\partial}{\partial\omega}M_1(h^{\prime},\omega)\right]\right|_{\omega=\epsilon(h^{\prime})}$ represents the lowest-order depletion of the Fermi Sea. The hole state $h'$ satisfies $|\vec{h'}|\leq k_F$. A representative value is $\kappa_2(h^{\prime}=0.75k_{F})=0.25$, indicating the non-negligible effect of this \emph{renormalization} term. In the present calculation, the self-energy is approximated up to third order as $\Sigma(k,\omega)\cong M_1(k,\omega)+M_2(k,\omega)+M_3(k,\omega)$. 

\subsection{The off-shell gap equation}
Using the nucleon self-energy, the gap equation with the off-shell propagator~\cite{fxh,gap} in the coupled channel reads 
\begin{align}
\begin{pmatrix}
\Delta_0 \\
\Delta_2
\end{pmatrix}(k)=\frac{1}{\pi}\int_0^\infty k^{\prime2}dk^{\prime}\left(\frac{1}{\pi}\int_0^\infty dE\mathrm{~Im}\Gamma(k^{\prime},E)\right)  \nonumber\\
\times
\begin{pmatrix}
V^{00} & V^{02} \\
V^{20} & V^{22}
\end{pmatrix}(k,k^{\prime})
\begin{pmatrix}
\Delta_0 \\
\Delta_2
\end{pmatrix}(k^{^{\prime}}),
\label{gap}
\end{align}
where $V^{LL'}(k,k')$ denotes the matrix elements of the realistic $NN$ interaction in the relevant coupled channels $(L,L'=0,2)$. The corresponding total gap is given by
$\Delta(k)=\sqrt{\Delta_{0}^{2}(k)+\Delta_{2}^{2}(k)}$. And the starting energy is redefined relative to the chemical potential, i.e., $E$ = $\omega-\mu$, for convenience. The kernel $\Gamma(k',E)$ is expressed as
\begin{align}
    \begin{aligned}
\Gamma(k,E) & =\mathcal{G}^{-}(k,-E)\mathcal{G}_s^{+}(k,E) \\
 & =[\mathcal{G}^{+}(k,E)^{-1}\mathcal{G}^{-}(k,-E)^{-1}+\Delta^2(k)]^{-1},
\end{aligned}
\end{align}
with the advanced and delayed Green functions given by 
\begin{align}
    \mathcal{G}^{+}(k,E)& =\Big[E-\xi_k-\Sigma(k,E)\Big]^{-1}\nonumber\\
    \mathcal{G}^{-}(k,-E )&=\Big[-E-\xi_k-\Sigma^{*}(k,-E)\Big]^{-1},
\end{align}
with $\xi_k=k^{2}/2m-\mu$. Accordingly, the full nucleon Green function in the BCS state is
\begin{align}
    \mathcal{G}_s^{+}(k,E)=&\Big[\mathcal{G}^+(k,E)^{-1}+\Delta^2(k)\mathcal{G}^-(k,-E)\Big]^{-1}.
\end{align}

The gap equation must be solved self-consistently with the density constraint due to the non-conservation of particle number in the BCS state. 
In the BCS state, the density is given by
\begin{align}
   \rho=4\sum_k\int_{-\infty}^0\frac{dE}{\pi}\mathrm{Im}\mathcal{G}_s^+(k,E),
    \label{rho}
\end{align}
where the factor of four accounts for spin and isospin degrees of freedom. Equation (\ref{gap}) should be solved self-consistently with this density constraint of Eq. (\ref{rho}) to obtain the pairing gap and the chemical potential.

\subsection{The spectral function and momentum distribution}
The nucleon spectral function in symmetric nuclear matter is obtained from the Lehmann representation of the Green function,
\begin{align}
    A(k,E)&=-2\mathrm{Im}\mathcal{G}^{+}(k,E)\nonumber\\
        &=\frac{-2\mathrm{Im}\Sigma(k,E)}{(E-k^2/2m-\mathrm{Re}\Sigma(k,E)+\mu)^2+\mathrm{Im}\Sigma(k,E)^2}.
\end{align}
Including the pairing gap, the spectral function reads
\begin{align}
    A_s(k,E)&=-2\mathrm{Im}\mathcal{G}_s^+(k,E)\nonumber\\
    & =-2\left\{\left(E+\xi_k+\mathrm{Re}\Sigma(k,-E)\right)^2\mathrm{Im}\Sigma(k,E)\right. \nonumber\\
 & \left. +\mathrm{Im}\Sigma(k,-E)\Delta^2(k)+\mathrm{Im}\Sigma(k,-E)^2\mathrm{Im}\Sigma(k,E)\right\} \nonumber\\
 &\Big/\left\{\big(\left(E-\!\xi_k-\!\mathrm{Re}\Sigma(k,E)\right)\left(E+\!\xi_k+\!\mathrm{Re}\Sigma(k,-E)\right)\right. \nonumber\\
&-\mathrm{Im}\Sigma(k,E)\mathrm{Im}\Sigma(k,-E)-\Delta^2(k)\big)^2 
\nonumber\\
&+\left(\mathrm{Im}\Sigma(k,E)(E+\xi_k+\mathrm{Re}\Sigma(k,-E)\right) \nonumber\\ 
 & \left.+\mathrm{Im}\Sigma(k,-E)(E-\xi_k-\mathrm{Re}\Sigma(k,E)))^2\right\}.
    \label{As}
\end{align}

The momentum distribution is obtained by integrating the spectral function over the full energy spectrum, weighted by the Fermi–Dirac distribution function:
\begin{align}
    n_s(k)=\int\frac{dE}{2\pi}A_s(k,E)f(E),
\end{align}
which reduces to the normal-state distribution when pairing is absent,
\begin{align}
    n(k)=\int\frac{dE}{2\pi}A(k,E)f(E).
\end{align}
At zero temperature, the integration is performed from $-\infty$ to $0$. Comparison of $n_s(k)$ and $n(k)$ allows separation of the pairing effects due to the tensor force from the contributions of SRCs to HMT.

\section{Results and discussions}

In the current calculation, only the realistic Argonne $V_{18}$ (Av18) two-body interaction is adopted as the pairing interaction, as the effect of the three-body forces on the $np$ pairing gap has been shown to be negligible~\cite{fxh}. This choice is also consistent with the self-energy calculation performed within the EBHF framework. In addition, the polarization corrections to the pairing interaction are omitted in the present work due to the inherent complexity of these effects, which remains an open problem.

\begin{figure}[H]
    \centering
\includegraphics[width=0.9\linewidth]{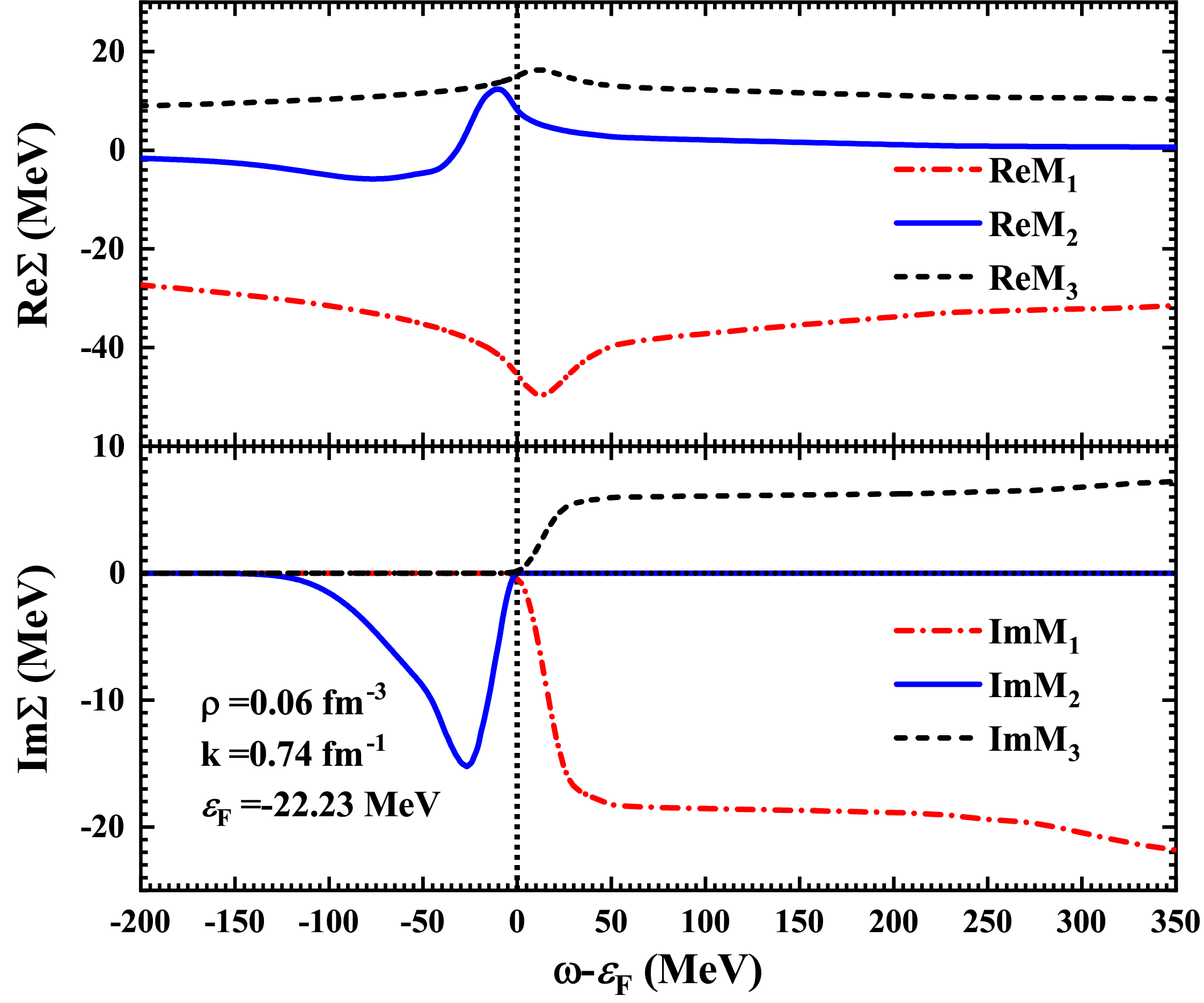} 
\caption{(Color online) The real parts (upper panel) and imaginary parts (lower panel) of the components $M_1$, $M_2$, and $M_3$ as functions of $\omega - \epsilon_F$ for $k = 0.74~\mathrm{fm}^{-1}$ at density $\rho = 0.06~\mathrm{fm}^{-3}$. $\epsilon_F$ denotes the single-particle energy at the Fermi surface in the BHF approximation. The short-dashed vertical line marks the position of the Fermi energy.}
\label{sigma}
\end{figure}

\begin{figure*}[t]
		\centering
		\subfigure{\includegraphics[width=0.43\linewidth]{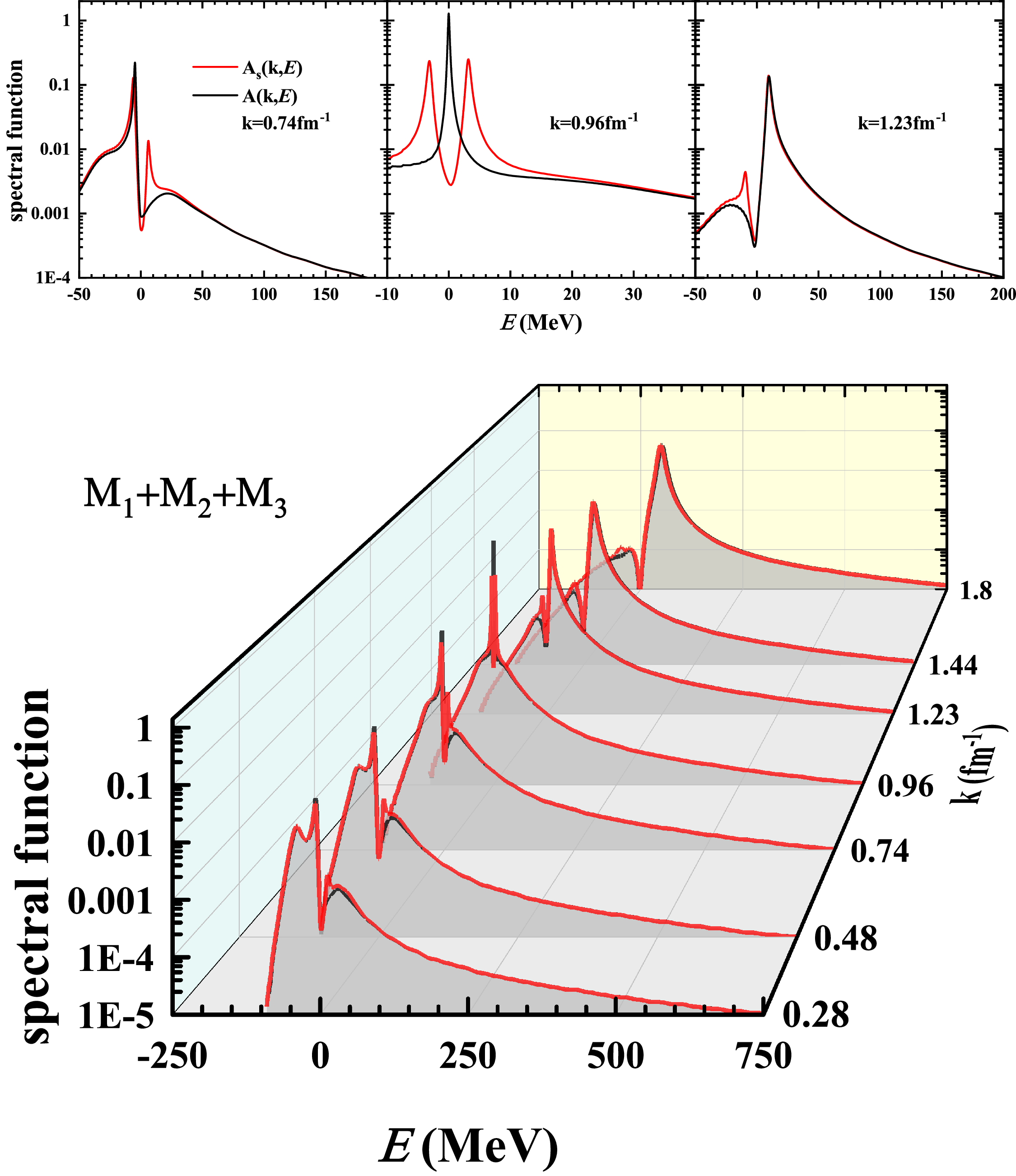}}
		\subfigure{\includegraphics[width=0.43\linewidth]{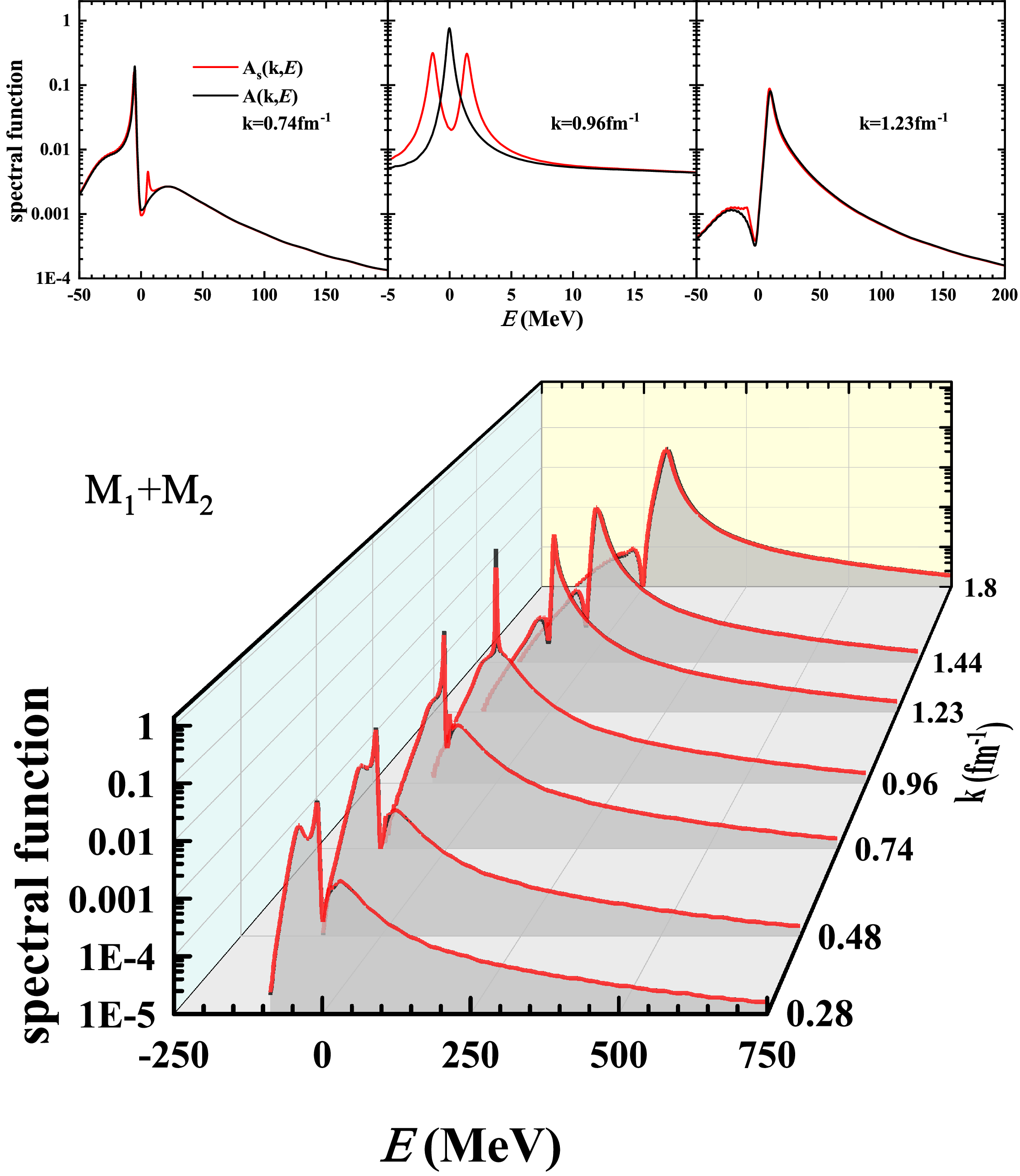}}
		\caption{(Color online) Nucleon spectral function in symmetric nuclear matter as a function of energy $E$ for various momentum $k$ at $\rho = 0.06~\mathrm{fm}^{-3}$. The inset panels show the spectral function for $k = 0.74$, $0.96$, and $1.23~\mathrm{fm}^{-1}$. The left (right) panel displays the results with (without) the third-order self-energy $M_{3}$. Results for both the BCS state and the normal state are shown for comparison.}
		\label{spectrum}
\end{figure*}

To illustrate the contributions of the different orders of the self-energy to the pairing gap and their effects on the momentum distribution, the real and imaginary parts of the individual components ($M_1$, $M_2$, and $M_3$) of the self-energy $\Sigma$ at a density of $\rho = 0.06~\mathrm{fm}^{-3}$ and fixed momentum $k = 0.74~\mathrm{fm}^{-1}$ are exhibited in Fig.~\ref{sigma}. The upper panel shows that the magnitude of $\mathrm{Re}M_3$ is significantly larger than that of $\mathrm{Re}M_2$, indicating that the $M_3$ contribution should be treated on the same footing as the second-order contribution $M_2$ for reliable predictions of the pairing gap. As discussed in the Formalism, $M_3$ accounts for the partial occupation of hole states below the Fermi surface induced by $NN$ correlations and usually contributes to the self-energy with a sign opposite to $M_1$. Since the self-energy modifies the density of states and hence the pairing gap, the inclusion of $M_3$, acting opposite to $M_1$, tends to enhance the pairing gap \cite{fxh}. 
$M_1$ is responsible for the depletion below the Fermi surface and exhibits a nonvanishing imaginary branch for $\omega > \epsilon_F$, whereas $M_2$ accounts for the high-momentum distribution above the Fermi surface with a nonvanishing imaginary part for $\omega < \epsilon_F$~\cite{Fermisea,M2}. Here $\epsilon_F$ denotes the single-particle energy at the Fermi surface in the BHF approximation. The two imaginary branches merge and vanish at $\omega = \epsilon_F$, as shown in the lower panel of Fig.~\ref{sigma}. This vanishing behavior ensures the reliability of the quasiparticle approximation near the Fermi surface. However, away from $\epsilon_F$, the imaginary part becomes comparable to the real part of the self-energy and should therefore be treated more accurately.

To provide a general understanding of the nucleon spectral functions in both the BCS and normal states, Fig.~\ref{spectrum} displays the nucleon spectral function of symmetric nuclear matter at the density of $\rho=0.06\ \mathrm{fm}^{-3}$ (corresponding to the Fermi momentum $k_{F}=0.9612 \ \mathrm{fm}^{-1}$), enabling a direct comparison between the two states as functions of the energy $E$ at different momenta. The red and black curves denote spectral functions in the BCS and normal states, $A_s(k,E)$ and $A(k,E)$, respectively. The left (right) panel shows results obtained with the self-energy calculated up to third (second) order. 

\begin{figure}[ht]
    \centering
    \includegraphics[width=0.9\linewidth]{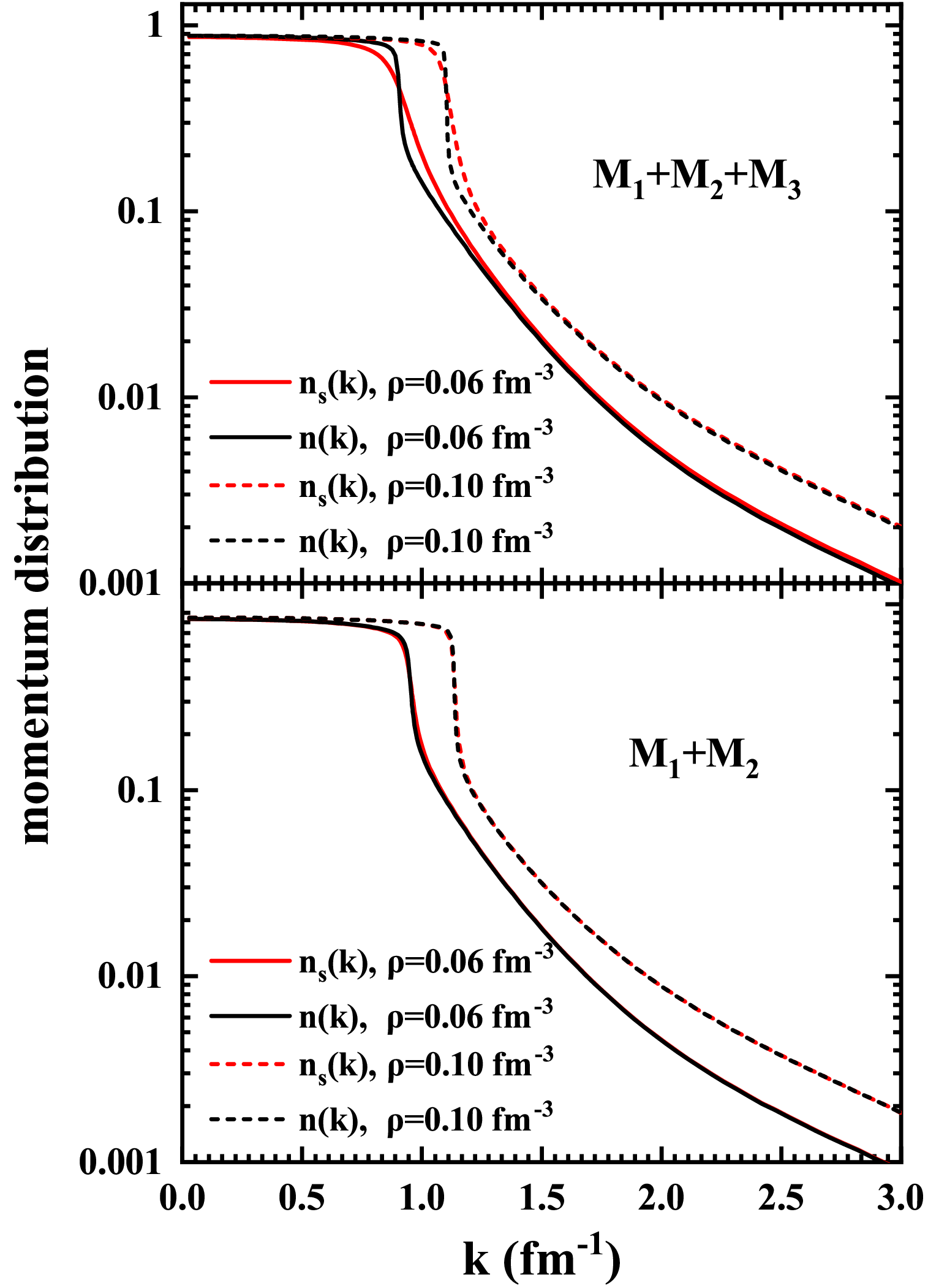}
    \caption{(Color online) Nucleon momentum distribution of the BCS state and normal state at $\rho=0.06\ \mathrm{fm}^{-3}$ and $\rho=0.10\ \mathrm{fm}^{-3}$. The upper panel shows the distributions including the third-order self-energy $M_3$, while the lower panel shows the distributions without $M_3$.}
    \label{nk}
\end{figure}

Away from the Fermi energy, i.e., $|E|\gg 0$, the spectral functions in the BCS and normal states nearly coincide, while noticeable differences appear only around the Fermi energy. This can be seen in the formula of $A_s(k,E)$, which shows that the influence of the pairing gap decreases rapidly as the energy departs from $E=0$. The insets illustrate the detailed structures of the spectral functions at three representative momenta, $k<k_{F}$, $k\approx k_{F}$, and $k>k_{F}$. For $k\approx k_{F}$, the normal-state spectral function can be approximated by $A(k,E)\approx Z_F\delta(E-E_{k})$. This $Z_F$ factor plays an important role in determining the transport properties of nuclear matter~\cite{SHANG2020}. Once pairing is included, the single-peaked structure characteristic of the normal state splits into two peaks corresponding to the time-reversed partner states forming a Cooper pair. For $k<k_{F}$, the two peaks of $A_s(k,E)$ become asymmetric due to the imaginary part of the self-energy. The peak below the Fermi energy nearly overlaps that in the normal state, while the peak above the Fermi energy becomes noticeably enhanced compared to $A(k,E)$, reflecting the increased depletion below the Fermi surface caused by pairing. For $k>k_{F}$, the situation is reversed as shown in the insets. These features are consistent with the results reported in Ref.~\cite{npa187}. In addition, a comparison between the left and right panels shows that the pairing gap obtained using the self-energy up to the third order ($M_3$) exhibits a more pronounced impact on the spectral function than that obtained from the second-order ($M_2$) calculation.

\begin{figure}[ht]
    \centering
    \includegraphics[width=0.9\linewidth]{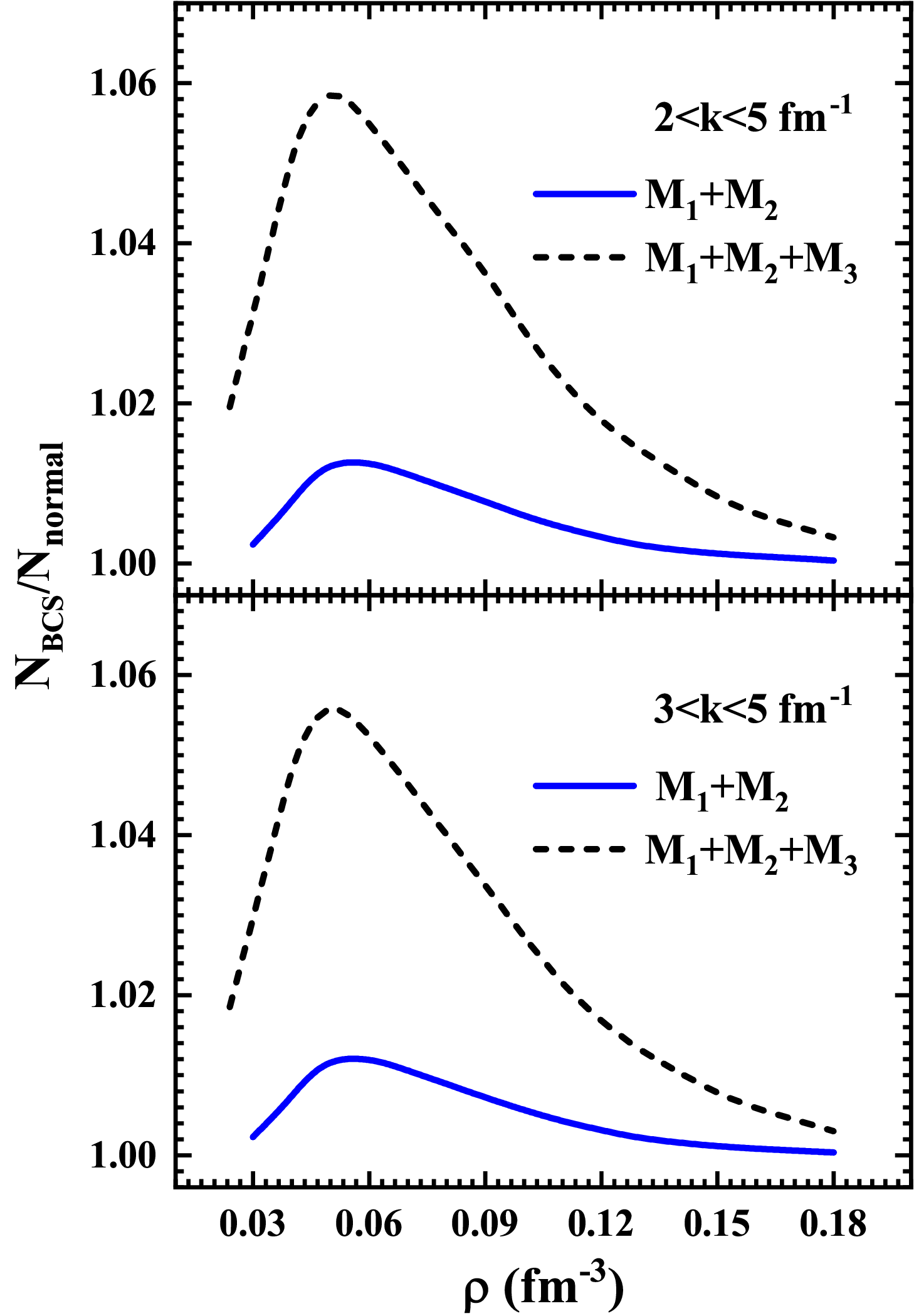}
    \caption{(Color online) The ratio of high-momentum nucleons between the BCS and normal states as a function of density, with and without $M_3$. The upper panel shows the ratio for nucleons with momenta in the range $2-5 \ \mathrm{fm}^{-1}$, while the lower panel shows the ratio for momenta in the range $3-5 \ \mathrm{fm}^{-1}$.}
    \label{nknk}
\end{figure}

The momentum distributions for the BCS and normal states are obtained by integrating the nucleon spectral function over energy. Fig.~\ref{nk} shows these distributions at densities of $0.06\ \mathrm{fm}^{-3}$ (solid lines) and $0.10\ \mathrm{fm}^{-3}$ (dashed lines), where the red and black curves represent the BCS and normal states, respectively. Even in the normal state, the momentum distribution reveals a depletion of the Fermi sea and a finite occupation of states above the Fermi surface, induced by $NN$ correlations. Once the pairing gap is included, as also indicated by the spectral function results, the momentum distribution exhibits the most significant deviation from the normal state near the Fermi momentum. Specifically, pairing enhances the depletion below the Fermi surface and increases the occupation above it. The resulting differences between $n_s(k)$ and $n(k)$ grow with the pairing gap size as illustrated in Fig.~\ref{nk}. It should be noted that the pairing gap decreases monotonously for densities $\rho > 0.06\ \mathrm{fm}^{-3}$~\cite{fxh}, and that the gap obtained from self-energy including contributions up to $M_3$ is significantly larger than that obtained from self-energy up to $M_2$. Moreover, as moving away from $k_F$, even for momenta $k\gg k_F$, the momentum distribution in the BCS state remains slightly higher than in the normal state, implying that the $np$ pairing may also enhance the HMT of the momentum distribution in symmetric nuclear matter.

To quantitatively isolate the contribution of pairing to HMT, we integrate the high-momentum components of the momentum distribution and define the HMT ratio as 
\begin{align}
    \rm {N_{BCS}/N_{normal}}=\frac{\frac{2}{\pi^2}\int_{k_a}^{k_\Lambda}n_s(k)k^2dk}{\frac{2}{\pi^2}\int_{k_a}^{k_\Lambda}n(k)k^2dk},
    \label{ratio} 
\end{align}
which characterizes the relative fraction of high-momentum nucleons in the BCS state compared with that in the normal state. In both the EBHF and off-shell BCS calculations, the maximum momentum cutoff is taken as $k_\Lambda=5\ \mathrm{fm}^{-1}$, ensuring convergence with respect to the momentum-space truncation. The resulting ratio is shown in Fig.~\ref{nknk} for two representative lower limits of integration, $k_a=2\ \mathrm{fm}^{-1}$ and $3\ \mathrm{fm}^{-1}$. Note that the tensor force dominates the HMT in nuclear matter in the region of $2-2.8\ \mathrm{fm}^{-1}$, whereas hard-core repulsion becomes dominant above $3\ \mathrm{fm}^{-1}$~\cite{2-2.8}. Therefore, these two choices of $k_a$ ensure that the corresponding high-momentum nucleons originate from SRCs.
Neutron–proton pairing enhances the population of high-momentum nucleons, resulting in a ratio greater than unity. The maximum value of $N_{BCS}/N_{normal}$ reaches approximately $1.06$ near $\rho=0.052\ \mathrm{fm}^{-3}$ when the \emph{renormalization} term $M_3$ is included. This result indicates that, in the HMT of symmetric nuclear matter, the contribution from the $np$ pairing might amount to about 6\% of that arising from SRCs. The density dependence of this HMT ratio with and without $M_3$ is qualitatively similar, although the magnitude is markedly reduced when $M_3$ is omitted. Furthermore, a comparison between the upper and lower panels shows that the results obtained with a lower integration limit of $k_a=3\ \mathrm{fm}^{-1}$ differ negligibly from those with $k_a=2\ \mathrm{fm}^{-1}$, expect a slight reduction. Such reduction arises from the fact that the difference between the BCS- and normal-state momentum distributions decreases with increasing momentum. Furthermore, because the HMT ratio reflects solely the effect of $np$ pairing, the small discrepancy between the result of $k_a=2\ \mathrm{fm}^{-1}$ and $k_a=3\ \mathrm{fm}^{-1}$ cannot reveal the distinct contributions to HMT originating from tensor force and hard-core repulsion.

\begin{figure}[H]
    \centering
\includegraphics[width=0.9\linewidth]{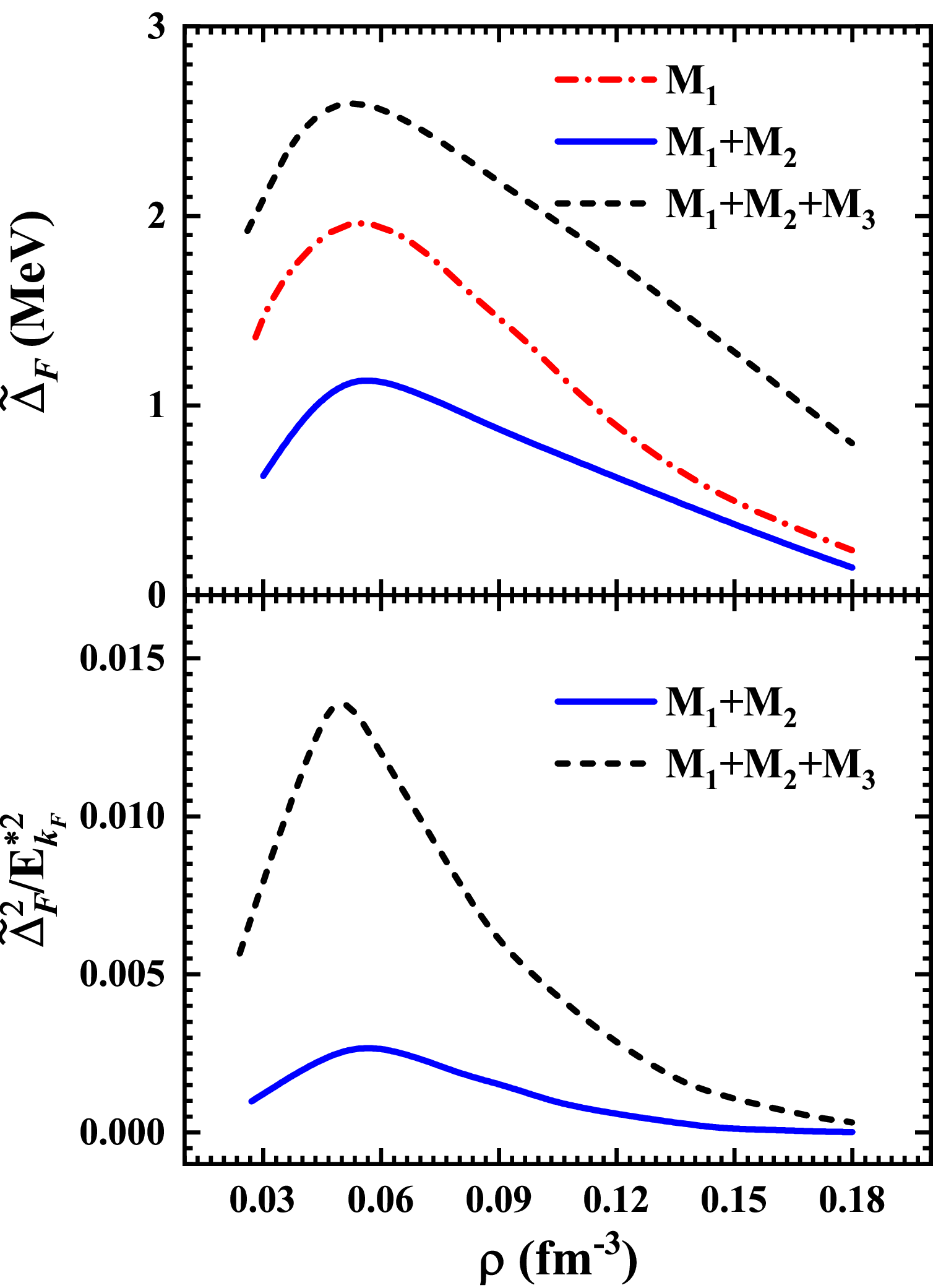} 
\caption{(Color online) The effective pairing gap [$\widetilde{\Delta}_F=Z_F\Delta(k_F)$] and the squared relative pairing gap $\widetilde{\Delta}_F^2/E_{k_F}^{*2}$ as a function of density, with different orders of self-energy.}
\label{delt}
\end{figure}

Essentially, the variation of the HMT ratio with density and for different self-energy treatments stems from the corresponding evolution of the pairing gap. Therefore, in the upper panel of Fig.~\ref{delt} we display the density dependence of the effective pairing gap [$\widetilde{\Delta}_F=Z_F\Delta(k_F)$] obtained with and without the inclusion of $M_3$. The density dependence of the pairing gap for $M_1+M_2$ and $M_1+M_2+M_3$ qualitatively follows the same trend as the HMT ratio. As discussed above, the inclusion of $M_3$ enhances the pairing gap, originating from the contribution of $M_3$ opposite to that of $M_1$.
To clarify more clearly the corrections to the pairing gap from different orders of the self-energy, we also show the result obtained with only $M_1$ included. It should be emphasized that, when only $M_1$ is retained, the absence of the imaginary part of the self-energy below $\epsilon_F$, as shown in Fig.~\ref{sigma}, leads to an inadequate description of the single-particle spectral functions $A_s$ and $A$. Consequently, the density constraint in Eq.~(\ref{rho}) is not well defined. To overcome this limitation, in the present calculation with only $M_1$ included we adopt the particle-number constraint as follow:
\begin{align}
    \rho&=4\sum_{k}n_s^{Z}(k)\nonumber\\&
    =4\sum_{k}\frac{Z_k}{2}\bigg\{1-\frac{\epsilon(k)-\mu}{\sqrt{\big[\epsilon(k)-\mu\big]^2+\big[Z_k\Delta(k)\big]^2}}\bigg\},
\end{align}\label{ns0}
with $Z^{-1}_k=1-\frac{\partial \Sigma(k,\omega)}{\partial\omega}\big|_{\omega=\epsilon(k)}$. While the pairing-gap equation is still given by Eq.~\ref{gap}.
As shown in Fig.~\ref{sigma}, although $\mathrm{Re}M_2$ is much smaller than $\mathrm{Re}M_1$ and thus may contribute only marginally to the modification of the pairing gap, the inclusion of $M_2$ introduces a sizable imaginary part that incorporates stronger $NN$ correlations. As a result, the pairing gap is quenched compared with the case including only $M_1$. In contrast, the inclusion of $M_3$ counteracts the $M_1$ contribution through both its real and imaginary parts, leading to an overall enhancement of the pairing gap. Consequently, the largest pairing gap is obtained with the $M_1+M_2+M_3$, followed by the $M_1$ result, while the smallest gap is found for $M_1+M_2$. However, the overall density dependence of the pairing gap remains similar in all cases, which may be mainly driven by the effective pairing interaction.

To further elucidate the effect of the pairing gap on the momentum distribution, $Z_k\Delta(k)$ in Eq.~\ref{ns0} can be approximated by its value at the Fermi surface, $\widetilde{\Delta}_{F}$. Under this approximation, the number of particles above the Fermi surface induced by the pairing gap can be well estimated as, 
\begin{align}
&\frac{2}{\pi^2}\int_{k_a}^{k_\Lambda}n_s^Z(k)k^2 dk\approx C\frac{\widetilde{\Delta}_{F}^2}{E_{k_F}^{*2}},
\end{align}\label{ns0}
valid for $\widetilde{\Delta}_{F}/E_{k_F}^* < 1$ and $k_a > 1.4k_F$, where $E_{k_F}^*=k_F^2/2m^*$ is the kinetic energy evaluated using the effective mass at $k_F$. With the inclusion of self-energy, the single-particle energy shift $\epsilon(k)-\mu$ in the BCS calculation is typically approximated by $(k^2-k_F^2)/2m^*$ (the effective mass approximation)~\cite{lombardo}. Accordingly, the relative pairing gap can be reasonably quantified by the ratio $\widetilde{\Delta}_{F}/E_{k_F}^{*}$. The effects of the self-energy on the pairing gap can be classified into two main contributions \cite{Lombardo2005}. First, the real part of the on-shell self-energy (single-particle potential) modifies the density of states at the Fermi surface, thereby affecting the pairing gap. This modification of the density of states is accounted for through the effective mass. Second, the imaginary part of the self-energy, together with the energy dependence of the real part, reflects $NN$ correlation effects (or equivalently, the $Z$ factor), which also significantly modify the gap. Therefore, in the present work, the effective-mass approximation provides an adequate description of the first contribution arising from the real part of the on-shell self-energy.

In the lower panel of Fig.~\ref{delt}, the density dependence of $\widetilde{\Delta}_{F}^2/E_{k_F}^{*2}$ for different orders of the self-energy is exhibited, with dashed (solid) lines corresponding to calculations with (without) $M_3$ contribution. The evolution of $\widetilde{\Delta}_{F}^2/E_{k_F}^{*2}$ closely follows that of the HMT ratio as a function of density. It should be noted that the difference between the BCS and normal states in the nucleon spectral function, which gives rise to the HMT ratio, is primarily governed by the squared relative pairing gap. Therefore, the density and self-energy dependence of the HMT ratio essentially originates from that of the relative pairing gap.


\section{Summary and Outlook}
In this work, we have investigated the HMT of nucleon momentum distributions in symmetric nuclear matter by consistently incorporating both SRCs and $np$ pairing effects within a combined framework of the extended EBHF approach and off-shell BCS theory. Specifically, nucleon spectral functions, momentum distributions, and the squared relative pairing gap $\widetilde{\Delta}_F^2/E_{k_F}^{*2}$ are calculated to quantify the contributions of $np$ pairing to the HMT. Because the EBHF self-energy includes SRCs through ladder resummations, the pairing effects are introduced on top of a realistic SRC background in a self-consistent manner.

Our results show that $np$ pairing significantly modifies nucleon spectral functions and momentum distributions around the Fermi momentum and produces a modest enhancement of the HMT in the BCS state compared to the normal state. The HMT ratio, used here to quantitatively characterize this incremental contribution of pairing, reaches a maximum value of about $1.06$. This indicates that the contribution of the $np$ pairing arising from the long-range tensor force to HMT amounts to approximately 6\% of that of SRCs at certain densities. The density and self-energy dependence of the HMT ratio closely follows that of the squared relative pairing gap, suggesting that $\widetilde{\Delta}_F^2/E_{k_F}^{*2}$ provides a qualitative measure of the influence of pairing on HMT. These findings confirm that SRCs are the primary origin of HMT, while the long-range tensor component associated with $np$ pairing may still generate observable effects under specific density conditions.

In the current study, only the realistic Av18 potential has been adopted. As is well known, the tensor force of the Av18 potential is relatively strong~\cite{fu2025}, it induces pronounced SRCs that suppress the $np$ pairing gap and its effects consequences. In contrast, in a new generation of high-precision $NN$ interactions, such as chiral interactions~\cite{chi1,chi2,chi3}, the tensor component is comparatively weaker~\cite{yin2023}, and therefore the interplay between SRCs and pairing may be altered. Within the chiral framework, the varying regulator cutoff and the chiral order further enable a systematic investigation of the competition between the short- and long-range tensor components of the tensor force. In addition, polarization corrections~\cite{polar1s02006} to the pairing interaction have not been included in the present calculation. Incorporating these effects will be an important issue for future work aiming to achieve a more comprehensive understanding of the contribution of $np$ pairing to HMT.

\section{Acknowledgments}
This work was supported by the National Natural Science Foundation of China under Grant No. 12375117, CAS Project for Young Scientists in Basic Research YSBR-088, the Youth Innovation Promotion Association of Chinese Academy of Sciences (Grant No. Y2021414), National Key R\&D Program of China No. 2024YFE0109802.

\bibliography{ref}

\end{document}